# The Falling Time of an Inverted Plane Pendulum

Milan Batista

University of Ljubljana, Faculty of Maritime Studies and Transportation

Pot pomorscakov 4, 6320 Portoroz, Slovenia, EU

milan.batista@fpp.uni-lj.si

Joze Peternelj

Division of Mathematics and Physics, Faculty of Civil and Geodetic Engineering,

University of Ljubljana, 1001 Ljubljana, Jamova 2, Slovenia, EU

joze.peternelj@fgg.uni-lj.si

(July 6, 2006)

**Abstract**

The article provides the formula for the calculation the falling time of inverted pendulum. The result is expressed in terms of elliptic integrals of first kind. The asymptotic formula for small initial inclination value is also provided.

**1 Introduction**

The problem of a pendulum is one of the oldest problem of classical mechanics and can be found in various textbooks ([5] pp 26, [9] pp 162-167, [9] pp 87-90). The main question of the problem is the period of free oscillations and the answer is given by

$$T = 2\sqrt{\frac{L}{g}}\int_0^{\theta_0} \frac{d\theta}{\sqrt{\cos\theta - \cos\theta_0}} = 2\sqrt{\frac{L}{g}}\int_0^{\theta_0}\frac{d\theta}{\sqrt{\sin^2\frac{\theta_0}{2} - \sin^2\frac{\theta}{2}}} = 4\sqrt{\frac{L}{g}}K\left(\sin\frac{\theta_0}{2}\right) \quad (1)$$

where *K* is complete elliptic integral of first kind (see Apendix). In the opposite to a pendulum in the problem of an inverted pendulum one of the questions is the falling time[1]. The purpose of the article is to answer this question.

**2 Solution**

Consider an inverted plane pendulum in the homogeneous gravitational field with acceleration *g*. The length of the pendulum is *L*, its mass is *m* and its moment of inertia is $J = mL^2$. The orientation (and hence the position) of the pendulum is determined by the inclination angle $\theta$ (Figure 1).

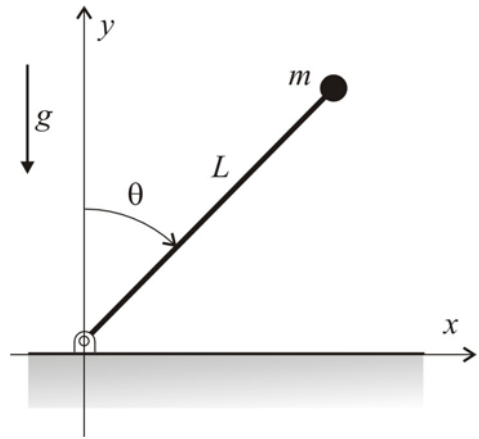

**Figure 1.** Pendulum variables

The dynamical equations of motion of the pendulum are reduced to the angular momentum equation

$$J \frac{d\omega}{dt} = mgL \sin\theta \qquad (2)$$

where $\omega$ is the angular velocity defined by

---

[1] This article does not enter into the questions connected with an inverted pendulum with horizontally or vertically moving point of suspension (upward-driven pendulum) which is one the famous stability and controlling problem ([2], [3])

$$\omega \equiv \frac{d\theta}{dt} \tag{3}$$

Taking into account that $\frac{d\omega}{dt} = \frac{d\omega}{d\theta}\frac{d\theta}{dt} = \frac{d}{d\theta}\left(\frac{\omega^2}{2}\right)$ equation (2) can be written in the form $\frac{d}{d\theta}\left(\frac{J\omega^2}{2}\right) = mgL\sin\theta$. This is ordinary differential equation with separable variables. Integration of the equation gives

$$E = \frac{J\omega^2}{2} + mgL\cos\theta \tag{4}$$

where $E$ is total mechanical energy of the pendulum which can be determinate by initial conditions $\theta_0 = \theta(0)$ and $\omega_0 = \omega(0)$

$$E = \frac{J\omega_0^2}{2} + mgL\cos\theta_0 \tag{5}$$

From (4) and (5) the angular velocity is

$$\omega(\theta) = \pm\sqrt{\frac{2mgL}{J}\left(\frac{E}{mgL} - \cos\theta\right)} = \pm\sqrt{\omega_0^2 + \frac{2g}{L}(\cos\theta_0 - \cos\theta)} \tag{6}$$

From (3) the differential equation for the time is $dt = d\theta/\omega(\theta)$. By combining this with (6) and integrate from the pendulum initial position $\theta_0$ to the final position $\pi/2$ (when pendulum hits the ground) gives the falling time of the pendulum

$$T(\theta_0, \omega_0) = \sqrt{\frac{L}{2g}} \int_{\theta_0}^{\pi/2} \frac{d\theta}{\sqrt{\frac{\omega_0^2 L}{2g} + \cos\theta_0 - \cos\theta}} \tag{7}$$

The rest of the article is devoted to evaluation of the integral in expression (7).

**3 Evaluation of the integral**

Consider the integral

$$I(\theta_0, a) = \int_{\theta_0}^{\pi/2} \frac{d\theta}{\sqrt{2a^2 + \cos\theta_0 - \cos\theta}} \qquad (8)$$

where for the case (7)

$$a \equiv \frac{\omega_0}{2}\sqrt{\frac{L}{g}} \qquad (9)$$

Before proceed note that the integral (8) is similar to the integral 2.571(5) ([4] pp 176)

$$\int \frac{dx}{\sqrt{b - c\cos x}} = \frac{2}{\sqrt{b+c}} F\left(\arcsin\sqrt{\frac{(b+c)(1-\cos x)}{2(b-c\cos x)}}, \sqrt{\frac{2c}{b+c}}\right) \qquad (10)$$

where $b > c > 0$, $0 \leq x \leq \pi$ and

$$F\left(\alpha, \frac{1}{p}\right) = p\int \frac{dx}{\sqrt{1 - p^2 \sin^2 x}} \qquad [p^2 > 1] \qquad (11)$$

Comparing (8) and (10) one finds that $b = 2a^2 + \cos\theta_0$ and $c = 1$. In the case $b < c$ (10) can not be directly applied to (8) so in the article instead of transforming (10), the required transformations will be done directly to the integral (8).

The integral (8) will be transformed to the elliptic integral in two steps. First, by means of trigonometric identity $\cos\theta \equiv 2\cos^2\frac{\theta}{2} - 1$ the integral (8) is transformed into

$$I(\theta_0, a) = \int_{\theta_0}^{\pi/2} \frac{d\theta}{\sqrt{2a^2 + 2\cos^2(\theta_0/2) - 2\cos^2(\theta/2)}}$$
$$= \frac{1}{\sqrt{2}k} \int_{\theta_0}^{\pi/2} \frac{d\theta}{\sqrt{1 - k^{-2}\cos^2(\theta/2)}} \qquad (12)$$

where

$$k = k(\theta_0, a) \equiv \sqrt{a^2 + \cos^2(\theta_0/2)} \qquad (13)$$

Depends on value of *k* three cases will be considered.

Case $k < 1$

In this case, from (13), one have $|a| < \sin(\theta_0/2)$. By defining the new variable as

$$\sin\varphi = \frac{\cos(\theta/2)}{k} \qquad (14)$$

integral (12) is future reduce into the integral

$$I(\theta_0, a) = \sqrt{2} \int_{\arcsin\frac{\sqrt{2}}{2k}}^{\arcsin\frac{\cos(\theta_0/2)}{k}} \frac{d\varphi}{\sqrt{1 - k^2 \sin^2\varphi}} = \sqrt{2} \int_{\frac{\sqrt{2}}{2k}}^{\frac{\cos(\theta_0/2)}{k}} \frac{du}{\sqrt{1 - u^2}\sqrt{1 - k^2 u^2}} \qquad (15)$$

where the second form is obtained by substitution $u = \sin\varphi$. From this one have

$$\boxed{\begin{aligned} I(\theta_0, a) &= \int_{\theta_0}^{\pi/2} \frac{d\theta}{\sqrt{2a^2 + \cos\theta_0 - \cos\theta}} \\ &= \sqrt{2}\left[ F\left(\frac{\cos(\theta_0/2)}{k}, k\right) - F\left(\frac{\sqrt{2}}{2k}, k\right) \right] \quad \left[k = \sqrt{a^2 + \cos^2(\theta_0/2)} < 1\right] \end{aligned}} \qquad (16)$$

where *F* is incomplete elliptic integral of the first kind (see Appendix). For special case $a = 0$ one have $k = \cos(\theta_0/2)$ so (16) become[2]

$$I(\theta_0) = \int_{\theta_0}^{\pi/2} \frac{d\theta}{\sqrt{\cos\theta_0 - \cos\theta}}$$
$$= \sqrt{2}\left[K(\cos(\theta_0/2)) - F\left(\frac{\sqrt{2}}{2\cos(\theta_0/2)}, \cos(\theta_0/2)\right)\right]$$

(17)

The two limit values for $I(\theta_0)$ are at $\theta_0 = 0$ and $\theta_0 = \pi/2$. In the first case the integral tends to infinity $I(0) = \infty$ and in the second its value is zero $I(\pi/2) = 0$. The graph of $I(\theta_0)$ calculated on the base of (17) is shown on Figure 1.

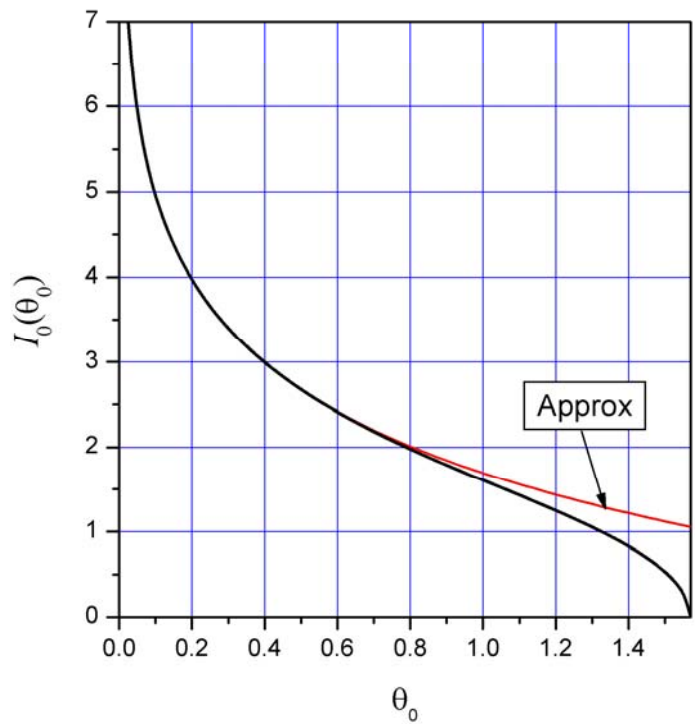

**Figure 1.** Graph of $I(\theta_0)$ (black) and its approximation (18) (red)

---

[2] The integral is also found in various other physical problems as for example in the calculation of the length of bent cantilever road ([5], pp 74)

The Table 1 provides compartment between results obtains by (17) and numerical evaluation of the integral by the Maple program with setting number of digits to 15 for various $\theta_0$.

**Table 1.** The values of $I(\theta_0)$ for selected $\theta_0$

| $\theta_0$ | Formula (17) | Maple numerical | Relative Error |
|---|---|---|---|
| $\pi/192$ | 7.510 737 00 | 7.510 737 00 | $0.7 \times 10^{-11}$ |
| $\pi/48$ | 5.550 960 41 | 5.550 961 59 | $0.2 \times 10^{-6}$ |
| $\pi/24$ | 4.572 120 47 | 4.572 120 47 | $0.3 \times 10^{-13}$ |
| $\pi/6$ | 2.609 754 96 | 2.609 754 96 | 0 |
| $\pi/3$ | 1.524 886 83 | 1.524 886 83 | $0.7 \times 10^{-14}$ |

For $\theta_0 \to 0$ one have the following limiting value ([1])

$$K\left(\cos(\theta_0/2)\right) \to \frac{1}{2}\ln\frac{16}{1-\cos^2(\theta_0/2)} = \frac{1}{2}\ln\frac{16}{\sin^2(\theta_0/2)} \approx \ln 8 - \ln\theta_0$$

Also for $\theta_0 \to 0$ one have

$$F\left(\frac{\sqrt{2}}{2\cos(\theta_0/2)}, \cos(\theta_0/2)\right) \to -\ln\left(\sqrt{2}-1\right)$$

so the asymptotic expansion for (17) is

$$\boxed{I(\theta_0) = \int_{\theta_0}^{\pi/2} \frac{d\theta}{\sqrt{\cos\theta_0 - \cos\theta}} \to \sqrt{2}\left[\ln 8\left(\sqrt{2}-1\right) - \ln\theta_0\right] \qquad [\theta_0 \to 0]} \qquad (18)$$

Some comportment of values calculated by (17) and (18) is given in Table 2. It is seen that approximation (18) far good even for larger initial inclination. For example for $\theta_0 = \pi/6 = 30^0$ is correct to three decimal places.

**Table 2.**

| $\theta_0$ | Formula (17) | Formula (18) | Relative Error |
|---|---|---|---|
| $\pi/3072 \approx 0.0010$ | 11.431 685 84 | 11.431 685 25 | $5.1 \times 10^{-8}$ |
| $\pi/384 \approx 0.008$ | 8.490 936 00 | 8.490 910 82 | $2.9 \times 10^{-6}$ |
| $\pi/96$ | 6.530 666 20 | 6.530 394 54 | $4.2 \times 10^{-5}$ |
| $\pi/24$ | 4.572 120 47 | 4.569 878 25 | $4.9 \times 10^{-4}$ |
| $\pi/6$ | 2.609 754 96 | 2.609 361 96 | $1.5 \times 10^{-4}$ |

*Note*

One can attempt to obtain the limit $\theta_0 \to 0$ by approximate $\cos\theta_0 \approx 1$ in (8). In this way the integral become

$$I(\theta_0) \approx \int_{\theta_0}^{\pi/2} \frac{d\theta}{\sqrt{1-\cos\theta}} = \frac{\sqrt{2}}{2} \int_{\theta_0}^{\pi/2} \frac{d\theta}{\sin(\theta/2)} = \sqrt{2} \ln\tan(\theta/4)\Big|_{\theta_0}^{\pi/2}$$
$$= \sqrt{2}\left(\ln\tan\frac{\pi}{8} - \ln\tan(\theta_0/4)\right) = \sqrt{2}\left[\ln 4(\sqrt{2}-1) - \ln\theta_0\right] + O(\theta_0^2) \quad (19)$$

The difference between (18) and (19) is $\ln 2 \approx 0.693$.

Case $k = 1$

In this case from (13)

$$a = \pm\sin(\theta_0/2) \quad (20)$$

and the integral (12) reduce to

$$I(\theta_0, a) = \frac{1}{\sqrt{2}} \int_{\theta_0}^{\pi/2} \frac{d\theta}{\sqrt{1 - \cos^2(\theta/2)}} = \frac{1}{\sqrt{2}} \int_{\theta_0}^{\pi/2} \frac{d\theta}{\sin(\theta/2)}$$
$$= \sqrt{2} \ln \tan \frac{\theta}{4} \bigg|_{\theta_0}^{\pi/2} = \sqrt{2} \left[ \ln\left(\sqrt{2} - 1\right) - \ln \tan \frac{\theta_0}{4} \right] \quad (21)$$

Case $k > 1$

As it is seen from (13) this case is only possible for $|a| > 0$ or more specifically for $|a| > \sin(\theta_0/2)$. The integral evaluation can be obtained directly from (15) by introducing new variable $w = k u$. By this (15) become

$$I(\theta_0, a) = \frac{\sqrt{2}}{k} \int_{\frac{\sqrt{2}}{2}}^{\cos(\theta_0/2)} \frac{dw}{\sqrt{1 - \frac{w^2}{k^2}} \sqrt{1 - w^2}} \quad (22)$$

From this

$$\boxed{\begin{aligned} I(\theta_0, a) &= \int_{\theta_0}^{\pi/2} \frac{d\theta}{\sqrt{2a^2 + \cos\theta_0 - \cos\theta}} \\ &= \frac{\sqrt{2}}{k} \left[ F\left(\cos(\theta_0/2), \frac{1}{k}\right) - F\left(\frac{\sqrt{2}}{2}, \frac{1}{k}\right) \right] \quad \left[ k = \sqrt{a^2 + \cos^2(\theta_0/2)} > 1 \right] \end{aligned}} \quad (23)$$

**4 Falling time**

Knowing $I(\theta_0, a)$ one can easily calculate the falling time of a pendulum by (7). For example let $L = 0.15\,\text{m}$, $g = 9.8\,\text{m/s}^2$ and $\omega_0 = 0$. Then one have $\sqrt{\frac{L}{2g}} \simeq 0.0875$ so the falling time for initial inclination of for example $\theta_0 = \pi/180 = 1^0$ is $T = 0.6490\,\text{s}$.

For $\theta_0 = 0$ the falling time is infinite $T = \infty$ i.e. the pendulum is in instable equilibrium position. This can be easely seen from (5). Namely the potential energy of the pendulum is $U = mgL\cos\theta$. From equilibrium condition $\frac{dU}{d\theta} = -mgL\sin\theta = 0$ one obtain $\theta = 0$ and since $\frac{d^2U}{d\theta^2} = -mgL\cos\theta$ it follows that since $\left(\frac{d^2U}{d\theta^2}\right)_{\theta=0} < 0$ the potential energy has maximum so the state is instable.

Maybe it is interesting to calculate the falling time when initial values are set according to quantum mechanics uncertainty principle. This principle state that $\Delta x \Delta p \geq \hbar/2$ where $\Delta x$ is position, $\Delta p$ linear momentum and $\hbar = 1.05457 \times 10^{-34}$ Js is the Planck constant divided by $2\pi$ ([7] pp 1002). In the case of inverted pendulum which is in the initial vertical position one have $\Delta x = L\sin\theta_0$ and $\Delta p = mL\omega_0$ so uncertainty principle transform to

$$\omega_0 \sin\theta_0 \geq \frac{\hbar}{2mL^2} \tag{24}$$

Taking the case $k = 1$ one obtain from (9) and (13)

$$\omega_0 = 2\sin\frac{\theta_0}{2}\sqrt{\frac{g}{L}} \tag{25}$$

Substituting this into (24) and taking into account that inclination angle is very small i.e. $\theta_0 \ll 1$ yield

$$\theta_0 \geq \sqrt{\frac{\hbar}{2mL^2}}\sqrt{\frac{L}{g}}$$

Now, by (21) the falling time is in this case

$$T \approx \sqrt{\frac{L}{g}} \left[ \ln 4\left(\sqrt{2}-1\right) - \ln \sqrt{\frac{\hbar}{2mL^2}\sqrt{\frac{L}{g}}} \right] \qquad (26)$$

For example let $L = 0.1\,\text{m}$, $m = 0.001\,\text{kg}$ and $g = 9.8\,\text{m/s}^2$. For these data the falling time is $T \approx 3.57\,\text{s}$.

**References.**

**Appendix.** Elliptic integral of first kind ([1] pp 589-626, [10] pp 315-332)

The incomplete elliptic integral of first kind is defined as

$$F(x,k) = \int_0^x \frac{du}{\sqrt{1-u^2}\sqrt{1-k^2 u^2}} \qquad [0 < k < 1] \qquad (27)$$

,

The complete elliptic integral of first kind which is defined as

$$K(k) = \int_0^{\pi/2} \frac{d\varphi}{\sqrt{1-k^2 \sin^2 \varphi}} = \int_0^1 \frac{du}{\sqrt{1-u^2}\sqrt{1-k^2 u^2}} \qquad (28)$$